\documentclass[twocolumn,twocolappendix]{aastex63}
\usepackage{graphicx}

\newcommand{\kmsmpc}{km\,s$^{-1}$\,Mpc$^{-1}$}
\newcommand{\kmsMpc}{km\,s$^{-1}$\,Mpc$^{-1}$}
\newcommand{\kms}{km\,s$^{-1}$}
\newcommand{\ho}{\ensuremath{H_{\rm 0}}}
\newcommand{\lcdm}{$\Lambda$CDM}
\newcommand{\hst}{\textit{HST}}
\newcommand{\gaia}{\textit{Gaia}}
\newcommand{\jwst}{\textit{JWST}}
\newcommand\mM{\ensuremath{(m{-}M)}}
\newcommand\zacs{\ifmmode z_{850}\else$z_{850}$\fi}
\newcommand\gacs{\ifmmode g_{475}\else$g_{475}$\fi}
\newcommand\gminz{\ensuremath{(g{-}z)}}

\newcommand\gzacs{\ensuremath{(g_{475}{-}z_{850})}}
\newcommand\JminH{\ensuremath{(J{-}H)}}
\newcommand\JHwfc{\ensuremath{(J_{110}{-}H_{160})}}

\newcommand\sna{SNe\,Ia}

\newcommand\mbarj{\ensuremath{\overline m_{110}}}
\newcommand\Mbarj{\ensuremath{\overline M_{110}}}

\newcommand\Mbar{\ensuremath{\overline{M}}}

\newcommand\Mtrgbacs{\ensuremath{M_{814}^{\mathrm{TRGB}}}}
\newcommand\MtrgbI{\ensuremath{M_{I}^{\mathrm{TRGB}}}}
\newcommand\chinu{\ensuremath{\chi^2_\nu}}
\newcommand\tM{\hbox{2M{\tt++}}}

\defcitealias{Jen21}{J21}

\shorttitle{The Hubble Constant from IR SBF}
\shortauthors{Blakeslee, Jensen, Ma, Milne, \& Greene}

\begin{document}

\title{The Hubble Constant from Infrared Surface Brightness Fluctuation
  Distances\footnote{Based on observations with the NASA/ESA \textit{Hubble Space Telescope},
    obtained at  the Space Telescope Science Institute, which is operated by AURA, Inc., under NASA
    \hbox{contract NAS\,5-26555}. These observations come from GO Programs \#11711, \#11712,
    \#12450, \#14219, \#14654, \#14771, \#14804, \#15265, \#15329.}}

\author[0000-0002-5213-3548]{John P. Blakeslee}
\affiliation{Gemini Observatory and NSF’s NOIRLab,
  Tucson, AZ 85719, USA, \,{\rm john.blakeslee@noirlab.edu}}

\author[0000-0001-8762-8906]{Joseph B. Jensen}
\affil{Utah Valley University, 800 W.~University Parkway, MS 179, Orem, UT 84058, USA}

\author[0000-0002-4430-102X]{Chung-Pei Ma}
\affiliation{Department of Astronomy, University of California, Berkeley, CA 94720, USA}

\author[0000-0002-0370-157X]{Peter A. Milne}
\affiliation{University of Arizona, Steward Observatory, 933 N.\ Cherry Avenue, Tucson, AZ 85721, USA}

\author[0000-0002-5612-3427]{Jenny E. Greene}
\affiliation{Department of Astrophysical Sciences, Princeton University, Princeton, NJ 08544, USA}

\begin{abstract}
 We present a measurement of the Hubble constant \ho\ from surface brightness fluctuation
 (SBF) distances for 63 bright, mainly early-type galaxies out to 100 Mpc observed with
 the Wide Field Camera~3 Infrared Channel (WFC3/IR) on the \emph{Hubble Space Telescope}
 (\hst). The sample is drawn from several independent \hst\ imaging programs using
 the F110W (1.1\,$\micron$) bandpass of WFC3/IR.  The majority of galaxies are in
 the 50 to 80 Mpc range and come from the MASSIVE galaxy survey.
 The median statistical uncertainty on individual distance measurements is 4\%.
 We construct the Hubble diagram with these IR SBF distances and constrain \ho\ using
 {four} different treatments of the galaxy velocities.
 For the SBF zero point calibration, we use both the existing tie to Cepheid variables, updated
 for consistency with the latest determination of the distance to the Large Magellanic Cloud from
 detached eclipsing binaries, and a new tie to the tip of the red giant branch (TRGB)
 calibrated from the maser distance to NGC\,4258.
 These two SBF calibrations are consistent with each other and with theoretical predictions from
 stellar population models. 
 From a weighted average of the Cepheid and TRGB calibrations, we derive
 $H_0=73.3{\,\pm\,}0.7{\,\pm\,}2.4$~\kmsmpc, where the error bars reflect the statistical
 and systematic uncertainties.  
 This result accords well with recent measurements of \ho\ from Type~Ia supernovae, time delays in
 multiply lensed quasars, and water masers.  The systematic uncertainty could be reduced to
 below 2\% by calibrating the SBF method with precision TRGB distances for a statistical sample of massive
 early-type galaxies out to the Virgo cluster measured with the \textit{James Webb Space Telescope}.~~~\\
 
\end{abstract}


\section{Introduction}


Ever since Cook's first expedition to Tahiti to observe the transit of Venus in 1769
\citep{cook71}, astronomers have been going to great lengths to measure accurate
distances and to corroborate their results through multiple independent routes \citep[see][]{hogg47}.
\hbox{Distances} enable us to convert the observed properties of planets, stars, galaxies, 
black holes, and cosmic explosions into physical quantities. They reveal the structure of the local
supercluster, map the peculiar motions, and constrain the present-day expansion rate,
parameterized by the Hubble constant \ho.
The successful gauging of distances beyond our planet has been the key to understanding the Universe.

Increasingly precise distances measured from stellar parallaxes in
the Milky Way \citep[e.g.,][]{lindegren2016,bailerjones2018}, detached eclipsing binaries
(DEBs) in the Large Magellanic Cloud (LMC) \citep{Piet2019}, Cepheids in nearby spirals
\citep[e.g.,][]{macri2015,Riess2020}, and a variety of other methods reaching out into the
Hubble flow, have progressively pushed down the
reported uncertainties on the local expansion rate.  For instance, using Type~Ia supernovae
(\sna) tied to Cepheids, in turn calibrated by a combination of Galactic parallaxes, DEBs in
the LMC, and the maser distance to NGC\,4258, \citet{Riess2019} find $H_0 = 74.03\pm1.42$
\kmsMpc.  However, \citet{Freedman2020} conclude
$H_0 = 69.6\pm0.8\pm1.7$ \kmsMpc\ from a calibration of \sna\ via the tip of the red giant branch
(TRGB), assuming the DEB distance to the LMC.  These two studies, which report precise values
of \ho\ that differ by nearly 2\,$\sigma$, use the same first and third rungs in their
distance ladders but differ in the intermediate step (Cepheids versus TRGB).~~~  

Also using the LMC-based TRGB method to calibrate \sna\ but with a different treatment of the
extinction for the LMC calibration stars, \cite{Yuan2019} report $H_0=72.4\pm2.0$ \kmsMpc.
Other recent estimates include (omitting units for brevity)  $H_0 = 71.1\pm 1.9$ from the \sna-TRGB
method calibrated by the maser distance to NGC\,4258 \citep{Reid2019}, $73.3\pm1.8$ 
from time delays in gravitationally-lensed quasars
\citep[][see also \citealt{Harvey2020}]{Wong2020},
$73.9\pm3.0$ from geometrical distances to galaxies hosting water masers \citep{Pesce2020},
and $75.1\pm3.0$ from the Tully-Fisher method calibrated with Cepheids \citep{cf4tf}.
In general, most techniques for measuring distances in the relatively local universe appear
to be converging on a value of $\ho\approx73$~\kmsMpc.  For a more comprehensive review of
the latest results in this rapidly evolving field, see \citet{divalentino2020}.


In parallel, the exquisitely precise measurements of the cosmic microwave background (CMB) by
the \textit{Planck} mission result in a predicted value of $H_0 = 67.4 \pm 0.5$ \kmsMpc, assuming the
standard ``lambda cold dark matter'' ($\Lambda$CDM) cosmology \citep{Planck2018}.
This lower value of \ho\ is not unique to \textit{Planck}, however.
Measurements of baryon acoustic oscillations (BAO) in redshift surveys, which probe
the scale of the primordial density fluctuations at later times, in combination with either
other CMB data or constraints from Big Bang nucleosynthesis, imply $H_0\approx67\pm1$
\citep{Addison2018,philcox2020}.
In addition, an analysis of the ``inverse distance ladder'' of \sna\ calibrated 
from BAO gives $H_0=67.8\pm1.3$ \kmsMpc\ \citep{inverseSNa}.

A significant discrepancy between the locally measured \ho\ and the value implied by
primordial density fluctuations within the context of \lcdm\ may point to physics beyond the
standard paradigm \citep[e.g.,][]{Planck2018,poulin2019,knox2020}. Formally, this
discrepancy now exceeds 4$\sigma.$  
Additional evidence for required extensions to \lcdm\ comes from the CMB itself, which shows a
higher than expected lensing amplitude in its power spectrum \citep{crisis2020}.
Alternatively, it could be an indication of unidentified systematics, perhaps lingering coherently in
multiple distance estimation methods. For this reason, additional routes to \ho\ are worth exploring.


One promising route is the surface brightness fluctuation method \citep[SBF,][]{Tonry1988,Jensen1998}.
The modern SBF method \citep[see the review by][]{blak2012} has 
become an all-purpose precision distance indicator for galaxies that are too distant for TRGB
detection, lack young populations that would host Cepheids, and have no known supernovae or
the extremely rare water megamasers.  Further, unlike Cepheids, \sna, and masers, which all
require data spanning multiple epochs, the SBF measurements require only a single
observation of sufficient depth, along with color data to provide the stellar population
calibration. As a result, the method has been applied widely to study the 3-D
structure of the nearest galaxy clusters \citep{Mei2007,Blak2009,mik2018ngvs}, determine the
physical size of the ``shadow'' (and thus the mass) of the supermassive black hole
in M87 \citep{ehtVI}, measure the most precise
distance to the host galaxy of the first gravitational wave source with an optical counterpart
\citep[GW170817;][]{mik2018n4993}, study the satellite systems of nearby
galaxies \citep{carlsten2019}, and confirm the nature of dark matter deficient ultra-diffuse
galaxies \citep{vanDokkum2018,cohen2018,BCrnaas2018}.
However, the analysis is not trivial. For an early exposition of the many things
can go wrong with SBF measurements, see \citet{bat99}, but most of these problems are avoided
with the superb resolution and image stability provided by the \textit{Hubble Space Telescope} (\hst).

The SBF method has been used less frequently to constrain \ho.  In some cases it has
served as an intermediate rung on the distance ladder between Cepheids and another method, such
as Tully-Fisher, the fundamental plane, or \sna\ \citep{Tonry1997, Blak2002, Khetan2020}. In
other cases, the \ho\ analysis relied on significant velocity corrections derived from
reconstructions of the local gravity field \citep{Blak1999, Tonry2000, Blak2002}.  Rarely have
SBF studies directly probed the Hubble flow with multiple galaxies --- notable exceptions are
\citet{Jensen2001} with \hst/NICMOS and \citet{biscardi2008} with \hst/ACS (discussed in
Sec.~\ref{sec:disc} below).  However, with the benefit of the Wide-Field Camera~3 Infrared
Channel (WFC3/IR) on \hst, and the calibration provided by \citet{Jensen2015}, we can now
obtain reliable SBF distances well out into the Hubble flow in a single \hst\ orbit.
In this paper, we present the first measurement of \ho\ based on a large sample of galaxies
with WFC3/IR SBF distances reaching out to 100~Mpc.

\section{IR SBF Distances}

The SBF method measures the small-scale spatial variance, or ``fluctuations,'' in 
intensity due to the discrete nature of the stars that comprise a galaxy.  Because the
fluctuations are dominated by red giant stars in the early-type galaxies we
target, the SBF signal is stronger in the near-infrared where these stars are brightest
and the exposure times for measuring SBF distances can be  much less than
in the optical.  The brighter fluctuations, combined with the lower IR background and stable
image quality from space, have made it possible to measure robust SBF distances with WFC3/IR
out to at least 100~Mpc in one to two \emph{HST} orbits.  As a result, we have been able to
amass a large sample of high-quality distances reaching as far as the Coma cluster in a
relatively modest amount of time. The galaxy observations come from several different
programs but have similar characteristics, as we detail below.

\subsection{Observational Data}

Our sample comprises 63 galaxies with WFC3/IR imaging in the F110W filter that we have
used to measure SBF distances based on the calibration by \citet{Jensen2015}.  The majority
of the observations come from a program (GO-14219) to obtain SBF distances to all galaxies
in the MASSIVE survey priority sample \citep{Ma2014} out to 6000~\kms. MASSIVE is a
volume-limited survey to study the structure, stellar populations, internal dynamics, and
central black holes of the most massive early-type galaxies within 100~Mpc
\citep[e.g.,][]{greene2015,veale2018,ene2020,liepold2020}. Uncertainties in
supermassive black hole masses for nearby galaxies are often dominated by distance errors
\citep[][]{Kormendy2013}.  Recognizing this, \citet{mcconnell-ma2013} used updated distances
for 44 galaxies in their black hole mass compilation, 41 of which were SBF values.
\citet{Goullaud2018} present an analysis of the surface brightness profiles for the galaxies
targeted in GO-14219.

The 6000~\kms\ limit ($\sim\,$80~Mpc) for the initial WFC3/IR SBF sample was chosen as the
point where the typical peculiar velocities of 300~\kms\ drop below 5\% of the Hubble
velocity, meaning that the relative distance error from redshifts is comparable to the
expected SBF distance error.  In addition, we calculated that in F110W we could reach far
enough along the globular cluster luminosity function (GCLF) at this distance
to reduce the contamination in the SBF power spectrum to $\sim\,$10\% in only one orbit.
The ability to remove contamination from globular clusters is usually our
limiting factor, rather than the signal-to-noise of the SBF signal itself, and this level of
detection completeness means that the error in this correction will be $\lesssim\,$0.03~mag.

Another large fraction of the galaxies in our data set come from a program (GO-14654) to
measure SBF distances to host galaxies of well-studied \sna\ with the goal of investigating
possible luminosity differences among subgroups with different ultraviolet colors
\citep[e.g.,][]{milne2013,milne2015,brown2017,brown2019,foley2020}.
We selected early-type hosts, or in some cases
disk galaxies that appeared to have useful regions for the SBF measurement, based on the
imaging data that were available when we were designing the program.  These galaxies are
naturally less luminous than those in the MASSIVE survey. For three targets in this
program that were expected to be near or beyond 80~Mpc, we conservatively obtained two
orbits of WFC3/F110W integration to ensure adequate depth along the GCLF.

In a follow-up to GO-14219, Program GO-15265 obtained second-band WFC3/UVIS imaging of
a subset of the previously targeted MASSIVE galaxies to study the metallicity
distributions of their globular clusters.  As part of this program we were also awarded time for
single-orbit WFC3/IR F110W imaging of an additional six MASSIVE galaxies beyond our initial
6000~\kms\ limit. We have found that we can reliably characterize the
GCLFs, and thus correct for their residual contributions to the variance, in these very
luminous ellipticals despite the somewhat larger distances. Thus, we have been able to
measure the SBF distances for these six galaxies and include them in our sample as well.
To these we add a two-orbit WFC3/F110W observation of the cD galaxy in Coma from another of
our programs and single orbit observations of two galaxies for which we previously published
WFC3/F110W SBF distances.

The following list summarizes the \hst\ programs that contributed to the
current data~set. All of the data were reduced homogeneously by our team.
\begin{itemize}
    \item GO-11711: NGC\,4874, the cD galaxy in the Coma cluster (PI: J.~Blakeslee). 
    \item GO-12450: NGC\,3504 (PI: C.~Kochanek), published previously by \citet{Nguyen2020}.
    \item GO-14219: 35 early-type galaxies selected from MASSIVE Survey (PI: J.~Blakeslee).
    \item GO-14654: 19 mainly early-type host galaxies of \sna\ (PI: P.~Milne).
    \item GO-15265: 6 additional MASSIVE Survey galaxies (PI: J. Blakeslee).
    \item GO-14771, GO-14804, GO-15329: NGC\,4993, host of the binary neutron star
      merger that produced GW170817 (PIs: N.~Tanvir, A.~Levan, E.~Berger), published previously by \citet{mik2018n4993}.  
\end{itemize}

In addition, we also reprocessed the observations from GO-11712 (PI: Blakeslee)
that were used by \citet{Jensen2015} to calibrate the WFC3/IR SBF method. This reprocessing
was done in order to verify consistency between the calibration and target samples. Although
\citet{Jensen2015} presented SBF calibrations for both F110W and F160W, we use only F110W
data in the present study and apply the same analysis for all program galaxies. This enables
an extremely homogeneous and self-consistent set of distance measurements.

\subsection{SBF Measurements}

The SBF analysis has been described many times
\citep[e.g.,][]{tal90,Jensen1998,bat99,mik2005,Mei2005}. For these \hst\ WFC3/IR data, we
follow the procedure documented for the calibration sample by \cite{Jensen2015}. Details
specific to the present sample of 63 galaxies, including some refinements to the masking
process and consistency tests with the calibration sample, are presented in a companion
data paper \citep[][hereafter \citetalias{Jen21}]{Jen21}.
In brief, we model and subtract the mean galaxy surface
brightness, mask  contaminating sources (mainly globular clusters and background
galaxies) down to some completeness threshold, and measure the Fourier power spectrum of the
remaining image. The normalization of the power spectrum on the scale of the point spread
function (PSF) includes contributions from the stellar fluctuations and contaminating sources.
We fit and extrapolate the magnitude distribution of the  sources and use this
to calculate the background variance, which we subtract from the normalization of the PSF
component of the power spectrum to get the variance from the stellar fluctuations.
These  fluctuations are measured in a series of concentric annuli, normalized
by the local surface brightness, and converted into the SBF magnitude, labeled \mbarj\ for
the F110W bandpass. Applying a calibration for \Mbarj\ (see the following section) then gives the
distance modulus.
%


%
We adopted a photometric zero point of 26.822 AB mag for WFC3/F110W from the STScI
website\footnote{https://www.stsci.edu/hst/instrumentation/wfc3/data-analysis/photometric-calibration}
at the time our data were obtained and processed. In December 2020, this value was revised
by $-0.004$~mag based on \citet{wfc2020}. Although this would change the
numerical value of the \mbarj\ measurements tabulated in \citetalias{Jen21} by a small
amount, it would not change our distances, which are referenced to the WFC3/IR SBF
calibration of \citet[][see also \citealt{mik2018n4993}]{Jensen2015}, who used the
former photometric zero point. We correct the photometry for Galactic extinction according
to \citet{sf2011}, using the values provided for individual galaxies by the NASA
Extragalactic Database (NED). Following that work, we adopted a 10\% uncertainty on the
extinction correction and included it in quadrature in our photometric error estimates.

\subsection{Color Dependence}

The absolute SBF magnitude \Mbar\ in a given bandpass depends on stellar population
properties such as age, metallicity, alpha-element enhancement, etc.  To calibrate this
dependence, one generally uses broadband color as a distance-independent proxy for a galaxy's
stellar population \citep[e.g.,][]{Tonry1997,Jensen2003,bva2001,mik2007}. \citet{Jensen2015}
derived high-quality calibrations of the F110W SBF magnitude \Mbarj\ using data for 16
Virgo and Fornax cluster early-type galaxies.  For the sake of flexibility, the calibrations
were provided using both ACS \gzacs\ and WFC3 \JHwfc\ colors.

Because the broad-baseline optical color is more sensitive to metallicity, the slope of the
\Mbarj{-}\gzacs\ calibration is shallower, and therefore less susceptible to photometric
errors; thus, it is the preferred calibration.  However, for galaxies with large amounts of
foreground extinction \citep[e.g.,][]{mik2018n4993}, or lacking in optical data, the
\Mbarj{-}\JHwfc\ calibration can be used.  For the current sample, we adopt the calibration
based on optical color for nearly all galaxies, but instead of ACS colors (unavailable for most
of the sample) we use colors derived from Pan-STARRS images \citep{magnier2020,waters2020},
with sky estimation and object masking as in \citet{Jensen2015}.
The photometric transformation of Pan-STARRS \gminz\ to ACS \gzacs\ is described in detail by
\citetalias{Jen21}. We include the scatter in this transformation in our estimate of the
color error, which contributes to the uncertainty in the calibrated \Mbarj\ for each galaxy.

One galaxy in our sample (ESO\,125-G006) lacks Pan-STARRS data; for this, we use 2MASS \JminH\ color
transformed to \JHwfc. As with \gminz, we include the scatter  
in this transformation in the estimated \Mbarj\ error.  This galaxy also suffers the highest
amount of Galactic extinction in our sample, 20\% higher than the next highest, and more
than three~times the sample average. Although the color measurement error is amplified
by the steeper slope of the \JminH\ calibration, the extinction is much lower in the IR,
resulting in an error for \Mbarj\ only slightly larger than for the rest of the galaxies.
The value of \ho\ derived from ESO\,125-G006 is very close to the best-fit value for
the full sample;  there is no evidence for any systematic difference resulting from the
use of \JminH\ for the calibration.
Likewise, \citet{mik2018n4993} found closely consistent distances for the gravitational wave event
host NGC\,4993 using the two different calibrations; \citetalias{Jen21} provide further
tests illustrating the consistency.

Finally, we note that the dependence of \Mbarj\ on galaxy color has some intrinsic scatter
due to stellar population effects.  For example, galaxies with the same colors may have
slightly different  \Mbarj\ values because age and metallicity are not completely
degenerate in their effects on \Mbarj\ and broadband color.  In the $z$-band this intrinsic
scatter is 0.06~mag for red galaxies \citep{Blak2009}.  Although the
observed scatter in the \Mbarj\ calibration from \citet{Jensen2015} is consistent with
measurement error, suggesting a negligible contribution from intrinsic scatter, we
conservatively adopt the same 0.06~mag intrinsic scatter for F110W.
With this added in quadrature, our median distance modulus error is 0.083~mag, or $\sim\,$4\% in distance.

\subsection{SBF-Cepheid Zero Point}\label{sec:zpt}

The $I$-band SBF distance zero point was tied to Cepheids by \citet{Tonry2000} using six
spirals that had both SBF and Cepheid distances.  This
calibration was revised by $+$0.06~mag by \citet{Blak2002} using the final Key Project
Cepheid distances from \citet{Freedman2001}, which were based on an LMC distance modulus of
18.50~mag. Additional discussion of this distance zero 
point, including checks for consistency with \hst/ACS SBF distances, is given in Appendix~A
of \citet{Blak2010}.  For WFC3/IR, the F110W and F160W SBF zero points were determined by
\citet{Jensen2015} using 16 Virgo and Fornax cluster galaxies with previously measured
SBF distances by \citet{Blak2009}.
\citet{mik2018n4993} revised these zero points by $0.05\pm0.02$~mag following
improved PSF characterization resulting from extensive tests with a library of template
stars that were used to reanalyze the power spectra of the calibration sample.  We use
the same set of PSF templates \citepalias[see][for details]{Jen21} and 
therefore adopt this zero-point shift and its associated scatter.

The above Cepheid calibration assumed a distance modulus of 18.50~mag for the LMC.
Based on a sample of 20 DEBs, \citet{Piet2019} present an improved LMC distance 
of $49.59\pm0.55$ kpc (combining random and systematic error), or $0.023\pm0.024$~mag
less than the value used previously.  For consistency with other recent studies
\citep[e.g.,][]{Riess2019,Freedman2019,Freedman2020}, we also apply this shift in
zero point. The fully revised calibration is presented by \citetalias{Jen21}.

The systematic uncertainty in the \Mbarj\ zero point was estimated by \citet{mik2018n4993} to
be 0.10~mag, including contributions of 0.03~mag from the tie between WFC3/IR and optical SBF
distances, 0.08~mag from the tie between SBF and Cepheids, and 0.06~mag for the Cepheid zero
point, dominated by the uncertainty in the LMC distance.
With the improved precision of the revised LMC distance, the Cepheid zero-point error drops
to 0.028 mag \citep{Riess2019}, reducing the SBF zero-point uncertainty to 0.09~mag,
or 4.2\% in distance. This is the current limit of our precision in measuring
\ho\ using the Cepheid-based SBF calibration alone.

Stellar population models provide confidence in this zero
point. Comparisons between SBF predictions for evolved stellar populations and the empirical
zero point show agreement in optical bandpasses to better than the 0.09~mag uncertainty
\citep[e.g.,][]{Blak2010,mik2018ngvs,greco2020}, although there is more 
discrepancy among  models for bluer populations \citep[e.g.,][]{trujillo2019}. The models
are less constraining in their predictions for near-IR SBF magnitudes, but they encompass the
range of the observations \citep{Jensen2003,Jensen2015}. Empirically, the tie between WFC3/IR and ACS
SBF distances is very tight.~~ 

\subsection{SBF-TRGB Zero Point}\label{sec:trgb}

Constraints on the zero point can be improved using the TRGB method for red galaxies with
well-measured SBF distances.  \citet{cohen2018} show excellent agreement between SBF and
TRGB distances for a sample of 12 dwarf galaxies with blue colors observed by \hst.
However, obtaining an overlapping sample of distances for massive red ellipticals requires
reaching the Virgo cluster; there are few TRGB distances for early-type galaxies at this
distance.  Two exceptions are M60 (NGC\,4649) \citep{LeeJang2017} and M87
\citep{bird2010}. There is also a TRGB distance to the massive merger remnant NGC\,1316
(Arp~154) in Fornax, which has multiple independent SBF distances
\citep[e.g.,][]{Blak2009,mik2013,Jensen2015}. In the Appendix, we compare the SBF
and TRGB distances for these three galaxies using an absolute magnitude for the TRGB itself
based on the maser distance to NGC\,4258 so that it is fully independent of the LMC-based
Cepheid calibration.  Excluding the problematic calibrator NGC\,1316, we find
that the mean offset between the SBF and TRGB distances is $-0.01\pm0.08$ mag. In other
words, SBF distances would be very slightly longer if calibrated from the TRGB. Including
systematic effects, the uncertainty on the TRGB-based SBF zero point becomes 0.10~mag, or 4.6\%.
Full details are provided in the Appendix.

Clearly there is no significant difference between the Cepheid-based and TRGB-based SBF
zero points.  However, the complete independence of the two approaches reduces the final
systematic uncertainty on the jointly constrained zero point from $>\,$4\% for each
method independently to 3.1\% when combined.  Ultimately, we hope to anchor the SBF method using 
a much larger sample of giant ellipticals with TRGB distances tied to
\gaia\ parallaxes.  We return to this prospect in Sec.~\ref{sec:future}.

\section{Velocity Data\label{sec:vel}}

In determining \ho, accurate velocities are equally important as accurate distances.  As
mentioned above, the majority of our galaxies are massive ellipticals in groups and clusters,
including several Abell clusters \citep[see][]{Ma2014}.  Many of these systems have velocity
dispersions in excess of 500 \kms, which would add significant scatter to the Hubble diagram
if using individual velocities.
We therefore use the group associations and mean group velocities in the CMB frame from
\citet{Tully2015}, based on the 2MASS Redshift Survey \citep[2MRS,][]{2mrs}. The 2MRS is 97.6\%
complete to magnitude $K_s{\,\leq\,}11.75$~mag over 91\% of the sky.  The Tully group catalog
contains cross-matched identifications to the \citet{Crook2007} catalog that was used
by \citet{Ma2014} and based on an earlier version of the 2MRS limited to $K_s{\,\leq\,}11.25$~mag.

For galaxies not in identifiable groups, we use the 2MRS velocity data directly.
We also test our results for \ho\ in the following section using only the individual,
rather than group, velocities.  In all cases we use velocities referenced to the CMB frame
and downloaded through the Extragalactic Distance
Database\footnote{http://edd.ifa.hawaii.edu/} \citep[EDD,][]{edd}. In fitting for \ho, we use the
first-order cosmological corrections to the redshift \citep[e.g.,][]{wright2006} assuming
$\Omega_m{=\,}0.3$, $\Lambda{\,=\,}0.7$; the results change imperceptibly if we instead use the
best-fit Planck values  ($0.315,0.685$) for a flat universe.


In the following section, we also calculate \ho\ using the flow-corrected velocities
predicted by two different models. 
The first is the Bayesian linear flow model of \citet{Graziani2019}, based on the
\textit{CosmicFlows-3} (CF3) database described by \citet{cf3}.
To implement this CF3 flow model for our sample, we use the online distance-velocity
calculator\footnote{http://edd.ifa.hawaii.edu/CF3calculator/api.php} described by \citet{cf3calc}.
The model returns the distance expected from the observed velocity in the Local Group frame.
To convert this model distance into the desired flow-corrected recessional velocity (i.e.,
Hubble expansion velocity), one must multiply it by the value of Hubble constant most
consistent with the CF3 database.
%
The question of what value of \ho\ to ``take out'' of the CF3 model in order to convert to
expansion velocity is not entirely trivial.  This value is a function of the 
zero point used for \sna\ in the CF3 database, and in that sense is
independent of the \ho\ we derive from our distances when using this velocity model.
\citet{cf3} find $H_0=76.2$ \kmsMpc\ to be the best-fit value for the CF3 database. In
constructing the flow model, \citet{Graziani2019} assumed a fiducial $H_0=75$ \kmsMpc;
then as part of the modeling, derived a best-fit scale factor of $1.02\pm0.01$, i.e.,
$H_0=76.5\pm0.8$~\kmsMpc.  To calculate the flow-corrected velocities for this model, we
adopt the best-fit \ho\ given by Tully and test the effect of the $\pm1$\% scale factor
uncertainty on our result.  When using this model, the scaling does affect our final \ho,
as discussed below.

The second flow model is that of \citet{Carrick2015}, based on the density field
reconstructed from the \tM\ redshift compilation \citep{Lavaux2011}, which
combines the 2MRS with deeper redshift surveys over large fractions of the sky.
We again use the online velocity calculator\footnote{https://cosmicflows.iap.fr} provided
for the model. Unlike the CF3 model, the \tM\ model calculator requires the input velocity
to be in the CMB frame.
\smallskip\smallskip

\section{The Hubble Constant}

The 63 galaxies in our sample have SBF distances ranging from 19 to 99~Mpc, were selected
in a variety of ways, and span a range of environments. Even homogeneously selected galaxies,
such as those in the MASSIVE sample, may exhibit diverse features that can
complicate the SBF analysis. In determining \ho, we use a weighted average in the
logarithm to ensure a symmetric treatment of the errors in distance modulus and
velocity; this is the equivalent of a single parameter fit minimizing $\chi^2$ for the sample.
To test the robustness of our results, we have performed the fits to \ho\ using numerous
different cuts and subsamples, as well as the {four} different approaches to the galaxy
recessional velocities described in the preceding section. The sections below present an
illustrative range of these fits before settling on a preferred value. All information
needed to reproduce and extend these tests is provided in \citetalias{Jen21}.

\begin{deluxetable}{lCccC}
\tablecaption{Hubble Constants for Various Selections\label{tab:hubble}}
\tabletypesize{\small}
\tablehead{\colhead{Selected sample\tablenotemark{a}} &
\colhead{ $v$\tablenotemark{b}} &
\colhead{$N_{\mathrm{gxy}}$} &
\colhead{\phm{m}$\chi^2_\nu$\phm{m}} &
\colhead{$H_0$} 
}
\startdata
 All galaxies       & grp & 63 & 1.19 &  73.53 \pm 0.66  \\ 
 Ellipticals        & grp & 45 & 1.02 &  73.52 \pm 0.74  \\ 
 All clean          & grp & 53 & 0.97 &  73.44 \pm 0.71  \\ 
 MASSIVE, clean     & grp & 37 & 1.16 &  73.86 \pm 0.82  \\ 
 $d>60$, clean      & grp & 34 & 0.88 &  73.33 \pm 0.82  \\ 
 $d<70$, clean      & grp & 33 & 0.88 &  74.08 \pm 0.96  \\ 
 $d<80$, clean      & grp & 46 & 0.96 &  72.78 \pm 0.77  \\ 
 SN Ia hosts, clean & grp & 20 & 0.68 &  73.31 \pm 1.26  \\[2pt] 
\hline
 All galaxies       & ind & 63 & 1.53 &  73.31 \pm 0.67  \\ 
 All clean          & ind & 53 & 0.95 &  73.27 \pm 0.73  \\ 
 MASSIVE, clean     & ind & 37 & 0.86 &  73.79 \pm 0.85  \\ 
 $d<80$, clean      & ind & 46 & 1.02 &  72.96 \pm 0.76  \\[2pt] 
\hline
 All galaxies       & cf3 & 63 & 1.14 &  73.32 \pm 0.71  \\ 
 All clean          & cf3 & 53 & 1.05 &  73.30 \pm 0.76  \\ 
 MASSIVE, clean     & cf3 & 37 & 1.16 &  73.62 \pm 0.88  \\ 
 $d<80$, clean      & cf3 & 46 & 1.07 &  72.67 \pm 0.83  \\[1pt] 
 All clean, $-$1\%\tablenotemark{c} & cf3 & 53 & 1.03 &  72.54 \pm 0.76  \\ 
 All clean, $+$1\%\tablenotemark{c} & cf3 & 53 & 1.06 &  74.07 \pm 0.77  \\[2pt] 
 \hline
All galaxies        & \tM & 63 & 0.99 &  73.90 \pm 0.65  \\ 
All clean           & \tM & 53 & 0.89 &  73.78 \pm 0.69  \\ 
Massive, clean      & \tM & 37 & 1.02 &  74.09 \pm 0.80  \\ 
 $d<80$, clean      & \tM & 46 & 0.94 &  73.42 \pm 0.75  \\ [1pt]
 \hline
\enddata
\vspace{-1pt}
\tablenotetext{a}{\footnotesize ``Ellipticals'' refers to morphological type $T\leq-3$; ``clean'' indicates galaxies with no discernible
dust or spiral structure; ``MASSIVE'' means limited to MASSIVE Survey galaxies.}
\vspace{-1pt}
\tablenotetext{b}{\footnotesize Velocities used for the fit: \textit{grp} for group-averaged;
  \textit{ind} for individual galaxy;  \textit{cf3}~for the flow model of \citet{Graziani2019};
  \tM\ for the flow model of \citet{Carrick2015}.}
\vspace{-1pt}
\tablenotetext{c}{\footnotesize Velocities from CF3 linear flow model rescaled by $\pm1$\%}
\vspace{-0.65cm}
\end{deluxetable}

\subsection{Group Velocities}


As a first approach, we use the group-averaged velocities in the CMB frame as described
above.  In doing this, it is necessary to adopt some value for the random dispersion of the
groups and isolated galaxies within the general velocity field. Estimates of this
small-scale peculiar velocity ``noise'' $\sigma_p$ range from 100 to 300~\kms, depending on
the sample considered \citep{masters2006,hoffman2015,Graziani2019}.  Here we assume
$\sigma_p=240$ \kms, which we add in quadrature with the generally small error in the group
mean velocity (taken as zero for isolated galaxies).  This results in a mean total velocity
error of 250~\kms, typical of many peculiar velocity surveys involving early-type galaxies
in groups \citep[e.g.,][]{zaroubi2001,hudson2004}. It corresponds to a 5.4\% velocity
error at the median distance of our sample.

\begin{figure*}
  \includegraphics[scale=0.263]{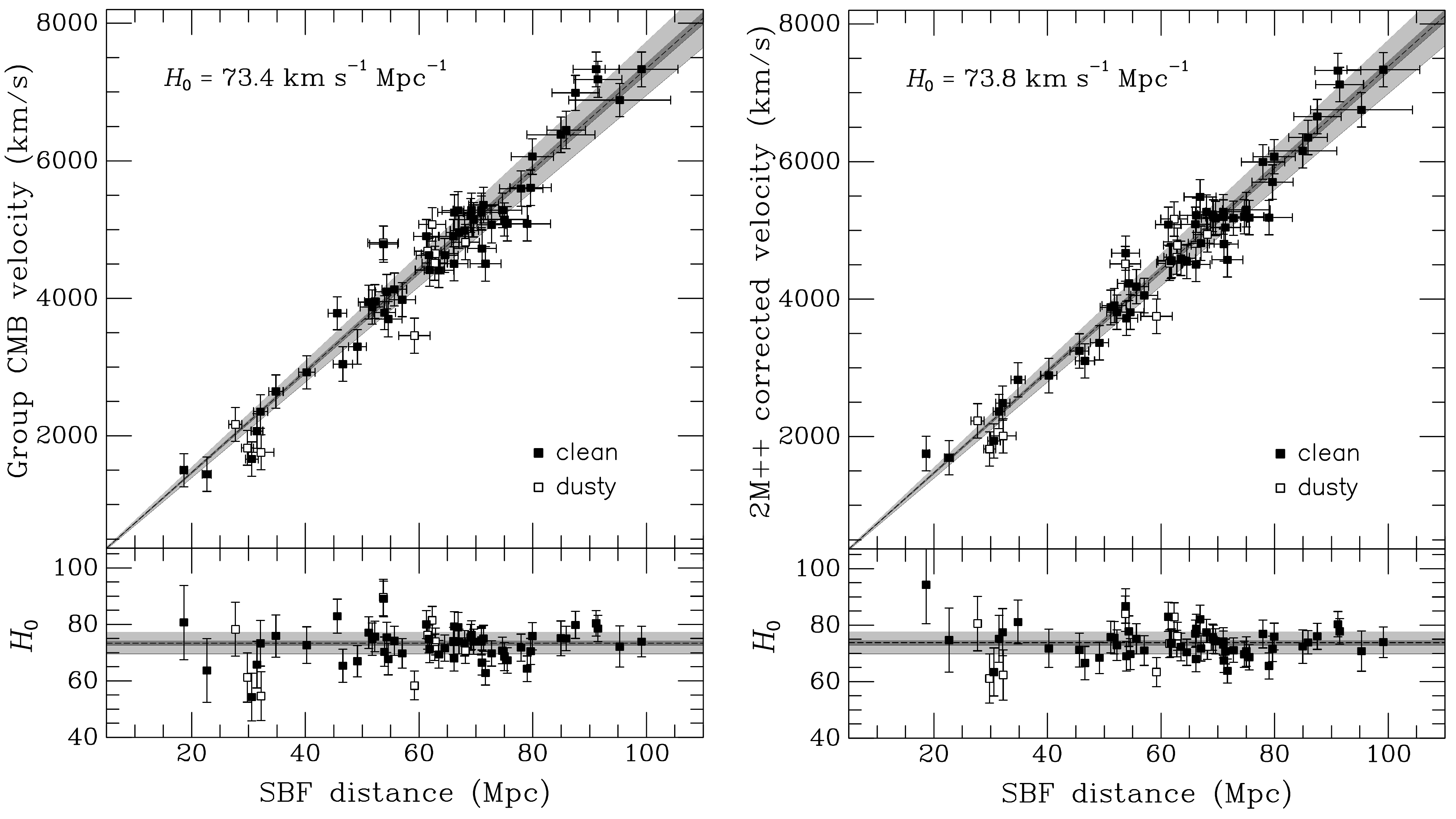}
  \vspace{-0.1cm}
  \caption{Left: Hubble diagram (top) and individual \ho\ values (bottom)
  for the Cepheid-calibrated WFC3/IR SBF distances and the galaxy
  group-averaged velocities in the CMB rest frame.  Solid symbols indicate ``clean''
  galaxies, for which no dust or 
  spiral structure is evident.\ 
  Open symbols indicate galaxies with obvious dust and/or spiral structure.
  The represented Hubble constant is the best-fitting value for the
  ``clean'' galaxy sample using these distances and velocities; the statistical and
  systematic error ranges are shown in dark and light gray, respectively.
  The plotted \ho\ error bars include both velocity and distance errors.
  Right:~Same as the plot on the left, except using the flow-corrected recessional
  velocities derived from the \tM\ density field analysis of \citet{Carrick2015}.
  The scatter is reduced by these flow-corrected velocities.
  Note that the distances would uniformly increase, and \ho\ decrease, by 0.3\% for the
  TRGB-based SBF calibration (see Appendix~\ref{sbf:trgb}).\\
\label{fig:hubble}}   
\end{figure*}

Table~\ref{tab:hubble} presents the results of many different trial fits of \ho, beginning
with the full sample of 63 galaxies with group-averaged velocities. For this case, we find
$H_0=73.53\pm0.66$ \kmsMpc, but with a $\chi^2$ per degree of freedom $\chinu=1.19$, which means
that either $\sigma_p$ or our distance errors are underestimated for at least a portion of
the sample.  Because morphological irregularities can cause problems for the SBF method, we
experimented by keeping only galaxies with morphological type $T\leq-3$ in the HyperLEDA
database \citep{leda}. This selects 45 galaxies classified as ellipticals or early-type S0s;
the fitted \ho\ is virtually identical, but with a significantly lower
$\chinu{=\,}1.02$. This is an acceptable fit for 44 degrees of freedom, indicating
that the errors are a reasonable description of the scatter for this sample.~~~ 

However, with our high-resolution \hst\ data, we can do better than catalogued types derived
from ground-based surveys.  In reducing the data, one of us (JBJ) made consistent note of
the presence of dust, spiral structure, bars, and shells.  This was done
prior to deriving the distances, so that knowledge of discrepant results could not bias the
classification.  Eleven galaxies show evidence of spiral structure or dust.
However, one of these is NGC\,4993, which has deep high-resolution ACS \gacs\ imaging that clearly
reveals the localized dust features, allowing them to be excised to give a clean area for
the SBF measurement.  A  distance for this galaxy was already published by
\citet{mik2018n4993}, and we keep it in our resulting sample of 53 ``clean'' galaxies
while rejecting the other ten.
Line~3 of Table~\ref{tab:hubble} shows that this sample gives  $H_0=73.44\pm0.71$ with
$\chinu = 0.97$, again indicating a good fit
(further cuts of galaxies with shells, bars, etc., do not yield additional improvement).
We also show results from a fit using only clean
galaxies that were selected as part of MASSIVE, a well-defined mass-selected
sample. Again, the value of \ho\ is consistent, but \chinu\ increases, likely because the
adopted $\sigma_p$ works well on average but slightly underestimates the small-scale
dispersion of galaxy groups embedded within rich environments.  A value of $\sigma_p=275$ \kms\
would give $\chinu = 1.0$ with negligible change to \ho.

Figure~\ref{fig:hubble} (left panel) presents the Hubble diagram combining our SBF distances
with the group-averaged velocities, along with the 
best-fit \ho\ value for the clean sample.  The figure also shows the values of
\ho\ derived from individual galaxies. The scatter decreases at larger distance as the
velocity errors become a smaller fraction of the recession velocity.  Overall the fit is
good, but by eye, there is a suggestion of small-scale coherent trends, with several
galaxies near $\sim\,75$~Mpc lying below the line and three near $\sim\,90$~Mpc scattering
above.  Features in the Hubble diagram can be associated with small-scale coherent
flows. There is an ongoing \hst/WFC3 SBF program exploring this issue in more detail.

Alternatively, if we were inadequately accounting for the effects of globular clusters in
the power spectra of the most distant galaxies, this could cause those galaxies to
scatter to high \ho.  We have explored a number of cuts in distance for our clean sample; the
table shows three examples: excluding galaxies within 60 Mpc, excluding those beyond 70 Mpc, or
beyond 80~Mpc.  In all these cases, $\chinu < 1$.  The nearer galaxies have little
leverage on \ho, and removing them has little effect. Changing the upper
distance limit can change \ho\ by up to $\pm1$\%, but this is still within the statistical error.

Finally, we also show the results from a fit using only clean galaxies that are identified
as \sna\ hosts. These galaxies range in distance from 19 to 91~Mpc, with the few most distant ones
having two orbits of integration.  The best-fit \ho\ is  similar to that derived for the full
clean sample, but the statistical uncertainty is much larger because the fit includes only
20~galaxies.


\subsection{Individual Velocities}

We performed a second set of fits using individual galaxy velocities from the 2MRS
referenced to the CMB frame.  However, we know that two-thirds of our sample consists of
massive ellipticals that preferentially reside in rich groups and clusters that can have velocity
dispersions up to 900~\kms.  Thus, the assumption of a fixed random velocity error is
insufficient. The \citet{Tully2015} group catalog provides line-of-sight velocity
dispersions for all groups, and these can be used as estimates of the random velocity errors 
for the member galaxies (neglecting the random motion of the groups themselves). We adopt
this approach: using the individual galaxy velocities with the errors taken to be the
dispersions of the parent groups.  For galaxies not in groups, we adopt an error of 150~\kms,
consistent with the quoted velocity field error in isolated regions from
\citet{Graziani2019}.

The second section of Table~\ref{tab:hubble} shows example fits using this approach.
Interestingly, the subsample of clean MASSIVE galaxies now gives the lowest \chinu.  This
is likely because the most massive members of groups and clusters tend to deviate from the
mean by less than one standard deviation, although there are exceptions in our sample
such as NGC\,1272 in Perseus, which has a velocity 1600~\kms\ below the cluster mean.
Overall, the \ho\ values from these fits track those from the group velocity fits
to within $\sim0.2$~\kmsMpc.

\subsection{Flow-Corrected Model Velocities\label{sec:flow}}
The lines labeled as ``\textit{cf3}'' in the second column
of {Table~\ref{tab:hubble}} show results obtained using the flow-corrected 
recessional velocities from the CF3 linear velocity model of \citet{Graziani2019}, implemented as
described in Sec.~\ref{sec:vel}.  For the velocity error in this case, we use the model's
best-fit nonlinear dispersion value of 280~\kms, which represents the unmodeled small-scale
motion in the velocity field.  The results of these tests are very similar to those found
using the individual CMB-frame velocities.  The \chinu\ values are a bit higher in most
cases, but there is no significant change in \ho.

We perform two other fits for this model, varying the scale factor of the velocity field within
the reported $\pm1$\% uncertainty. Unsurprisingly, this changes the \ho\ we
derive by $\pm1$\% as well.  Since the result given by the baseline scale factor agrees well
with that found using the uncorrected CMB-frame velocities, this  change in
\ho\ would represent a net inward or outward flow of our sample with respect to
the comoving volume (a small Hubble bubble or sink-hole).  Until we have more complete
high-precision distance mapping of the local volume extending beyond 100~Mpc, we take this $\pm1$\%
as the systematic uncertainty in the velocity scale.

Finally, the fits in  Table~\ref{tab:hubble} identified as ``2M\texttt{++}'' use the
flow-corrected recessional velocities from the density-based model of
\citet{Carrick2015}, also described in Sec.\,\ref{sec:vel}.
As for the group velocity fits above, we again use $\sigma_p=240$ \kms.
We find that the corrected velocities from this model result in significant improvements in
\chinu\ with respect to the other sets of velocities.  For instance, the unreduced $\chi^2$
for the full sample drops by $\Delta\chi^2>12$ compared to the fit using the group-averaged
velocities. The resulting \ho\ values are $\sim\,0.4$ \kmsMpc\ higher, again consistent
within the errors.
The Hubble diagram in the right panel of Figure~\ref{fig:hubble} illustrates the slightly
higher value of \ho\ and reduced scatter when using the \tM\ model velocities.

\subsection{Final $H_0$}


For the preferred $H_0$ from our analysis, we adopt the result in Table~\ref{tab:hubble} for
the full ``clean'' sample using the group velocities.\  
Although the \tM\ model velocities gave lower values of \chinu,
we suspect that the model-independent group velocities may be more robust against systematic
errors in scaling. We note that same approach was used by \citet{Pesce2020} in deriving
\ho\ from their maser distances, for which the \tM\ model similarly gave the best fit.
Readers who prefer the \ho\ result based on the \tM\ model should increase the values
given in this section by 0.5\%.%
Thus, we adopt $\ho=73.44\pm0.71$ \kmsMpc, statistical error only,
for our Cepheid-calibrated value of $H_0$.
The revised Cepheid-based SBF calibration has a systematic uncertainty of
0.09~mag, or 4.2\% in distance (\hbox{Sec.\,\ref{sec:zpt}}), translating to
$\pm\,3.11$~\kmsMpc\ for \ho.   

In Appendix~\ref{sbf:trgb}, we find that the TRGB-based calibration differs by
$0.007{\,\pm\,}0.099$~mag from the Cepheid calibration,
in the sense that the SBF distances increase by $\sim\,$0.3\%.
The TRGB calibration therefore gives $\ho=73.20$~\kmsMpc, with a systematic
uncertainty of 4.7\%, or $\pm\,3.41$~\kmsMpc.
Averaging these two independent calibrations gives 73.33~\kmsMpc\ with a systematic error in
distance of 3.1\%. We combine this in quadrature with the
estimated 1\% systematic error in the velocity scaling (Sec.~\ref{sec:flow}) to give a total
systematic error of 3.3\%, or 2.41~\kmsMpc.
Our final result, to single decimal-point precision,
is then $H_0 = 73.3\pm0.7\pm2.4$~\kmsMpc.
This calculation is summarized in Table~\ref{tab:H0final}.

\begin{deluxetable}{lcccc}
\tablecaption{Final Hubble Constant and Errors \label{tab:H0final}}
\tabletypesize{\small}
\tablehead{
\colhead{SBF calibration} &
\colhead{\phn\phn$H_0$\tablenotemark{\small a}\phn} &
\colhead{\phn$\sigma_{\mathrm{stat}}$\tablenotemark{\small b}} &
\colhead{\phn$\sigma_{\mathrm{sys}}(d)$\tablenotemark{\small c}} &
\colhead{$\sigma_{\mathrm{sys}}(v)$\tablenotemark{\small d}} 
}
\startdata
 Cepheid    & 73.44 & 1.0\% & 4.1\% &  1.0\%  \\ 
 TRGB       & 73.20 & 1.0\% & 4.7\% &  1.0\%  \\[1pt] 
\hline
Average    & 73.33 & 1.0\% & 3.1\% &  1.0\%  \\[2pt]
\hline  \multicolumn{5}{c}{Final:\,
   $\; H_0 \,=\; 73.3 \pm 0.7 \pm 2.4$ \kmsMpc}  \\[2pt]
\enddata
\vspace{-1pt}
\tablenotetext{\small a}{\ho\ for ``clean'' galaxy sample with group velocities.}
\vspace{-2pt}
\tablenotetext{\small b}{Statistical error from \ho\ fit.}
\vspace{-2pt}
\tablenotetext{\small c}{Systematic uncertainty in distance calibration.}
\vspace{-2pt}
\tablenotetext{\small d}{Systematic uncertainty in velocity scaling.}
\vspace{-0.76cm}
\end{deluxetable}

\section{Context}
\label{sec:disc}

\citet{Verde2019} referenced a preliminary value of~\ho\ from this project
presented at a 2019 Kavli workshop \citep{blak-kavli2019}. The value presented here is 4\%
lower than the preliminary one. There have been several changes in our analysis since that workshop.
First, we have added nine more galaxies to our WFC3/IR sample, improving our
constraints.  Second, as discussed by \citetalias{Jen21}, we have improved our masking
technique for removing background sources. This has negligible effect for most
galaxies in our sample, but becomes marginally significant for the most distant galaxies
with only one orbit of integration; we explored this issue in our tests above by imposing
distance cutoffs on the sample.

However, the main reason for the decrease in \ho\ was our switch from the velocities given
by the \citet{Mould2000} parametric attractor model to those used here. As noted by
\citet{Tonry2000}, models that add massive attractors without compensating to ensure zero
net change to $\Omega_m$ can bias \ho\ high if the galaxies preferentially lie beyond
the attractors, as is the case here for the Virgo and Great Attractor components of the
Mould model. We note that \citet{Pesce2020} saw the same effect: their \ho\ increased by
4\% when they used the Mould model velocities.  Given the consistency of our results for
{four} different treatments of the recessional velocities,
including two recent flow models, we are confident that the
\ho\ presented here is far more robust than the earlier one.  

As discussed in the Introduction, many of the previous measurements of \ho\ using the SBF
method relied on other techniques to tie local SBF distances to the far-field Hubble flow.
A recent example is by \citet{Khetan2020}, who used a heterogeneous set of 24
ground-based and \hst\ SBF distances to calibrate
the \sna\ peak luminosity, and then derived \ho\ from 96 \sna\ at larger distances. That
study used essentially the same distance calibration as us, except for the adjustment we
made for the improved LMC distance.  Including this revision, the
\citet{Khetan2020} value becomes $H_0=71.1 \pm 2.4 \pm 3.4$ \kmsMpc\ (statistical and
systematic errors), which agrees with our
result to within the statistical error.  Of~course, since there are
significantly fewer Cepheid calibrators for the SBF method than for \sna, using SBF as an
intermediary will result in lower precision on \ho.  Nonetheless, 
SBF distances remain an extremely promising way of exploring possible
systematics in \sna\ properties; \hst\ program GO-14654 was predicated on this very  goal
and has provided nearly a third of our WFC3/IR SBF sample. Our results from this program
will be presented in future works (Garnavich et al., in prep.; Milne et al., in prep.).

Although ours is the first measurement of \ho\ from a statistical sample of WFC3/IR SBF 
distances, \citet{Jensen2001} previously derived $H_0=76\pm1.3\pm6$ \kmsmpc\ from
IR SBF distances for 16 galaxies ranging from 40 to 130~Mpc observed with NICMOS. When 
considering only the six most distant galaxies, their value was $72\pm2.3\pm6$
\kmsmpc. However, as discussed in that work, the NICMOS images suffered 
persistent residuals from cosmic rays that affected the image power spectra to
an uncertain degree.  Further, the NICMOS SBF calibration, explored in more detail
by \citet[][]{Jensen2003}, was not well constrained for stellar population effects,
and \citet{Blak2010} discuss a
possible systematic shift in the NICMOS distance moduli.  In contrast, the WFC3/IR SBF
calibration by \citet{Jensen2015}, with the small revisions discussed in Sec.~\ref{sec:zpt}
above, provides a firm foundation for the current \ho\ measurement.

Another direct SBF \ho\ measurement  was by \citet{biscardi2008} who analyzed
ACS/F814W images of four galaxies with distances of 55 to 110~Mpc. Interestingly, these
authors opted for a theoretical calibration from the stellar population models of
\citet{raimondo2009}.  Their result was $H_0 = 76\pm6\pm5$ \kmsMpc, where the second error
bar reflects their estimated uncertainty in the model calibration.  Further refinements to
the models, perhaps incorporating constraints from \textit{Gaia}, could make this a
competitive calibration route for SBF.

\citet{mik2018n4993} previously demonstrated the usefulness of single-orbit WFC3/IR
observations for precise SBF distances out to at least 40~Mpc and reported
a Hubble constant $H_0=71.9\pm7.1$ \kmsMpc\ (or $72.5{\,\pm\,}7.2$ with the revision to the LMC
distance). Because that sample consisted of only NGC\,4993, the
host galaxy of the GW170817 event, the total error was dominated by statistical effects,
including both the distance measurement error and the random error
in the peculiar velocity. With the present sample of over 60 galaxies, NGC\,4993 among
them, the statistical error in \ho\ has dropped below the 1\% level.


In the broader context, our result is consistent with recent values of
\ho\ from Cepheid-calibrated \sna\ and Tully-Fisher distances in the local universe, as
well as Cepheid-independent results from water masers and gravitationally lensed image delay times
(see references in the Introduction).  It disagrees with the predicted value from
\textit{Planck} for the \lcdm\ cosmology at the $2.3\sigma$ level.
Although our measurement is by far the most direct and precise value of \ho\ from SBF
to date, it could be improved substantially through a better constraint on the distance
zero point.  We conclude by suggesting one possibility for accomplishing this goal.

\vspace{2pt}

\section{Summary and Outlook}
\label{sec:future} 


We have presented a new measurement of the Hubble constant based on a homogeneous set of
63 WFC3/IR SBF distances extending out to 100~Mpc. The data~are provided in a companion
paper \citepalias{Jen21}.\ Two-thirds of our sample are luminous early-type galaxies
selected as part of the \hbox{MASSIVE} Survey; 
most of the rest were selected as hosts of recent \sna.
We have performed numerous tests, including {four} different treatments of the galaxy
velocities, to ensure that our result is robust within the quoted errors.
Although the \tM\ flow model gives the lowest scatter in the Hubble diagram, we use the
observed group-averaged velocities to retain model independence.
Averaging the results for the Cepheid and TRGB calibrations of SBF,
our~final result is $H_0 = 73.3\pm0.7\pm2.4$~\kmsMpc, where the error bars represent
the statistical and systematic uncertainties.  This agrees well with other recent 
measurements of \ho\ in the local universe.

Of the 63 galaxies in our WFC3/IR SBF data sample, 24 have hosted \sna\ with
high-quality light curve data.\  The intercomparison of these two precision distance
indicators enables detailed investigation of subtle effects in both methods
(Milne et al., in prep.).\ It may also provide a better absolute calibration of the
SBF zero point, since \sna\ have been tied to Cepheids through a larger number of
calibrating galaxies. Additionally, we have an ongoing WFC3/IR SNAP program that will
further improve our constraints on \ho\ and the velocity structure of the local universe.
At the same time, more extensive stellar population modeling in the near-infrared,
implementing new constraints on the luminosities and lifetimes of evolved stars in
metal-rich populations \citep[e.g.,][]{villaume2017,Pastorelli2020}, can help us
understand better the deviations of some galaxies from the IR SBF versus color relation,
perhaps leading to an improved distance method.


Although WFC3/IR has proven to be a powerful tool for SBF measurements, reaching 80 to
100~Mpc with \hbox{4-5}\% distance precision in a single \hst\ orbit, the systematic uncertainty
in the SBF zero point currently limits the  method's ability to achieve competitive
constraints on cosmology.  However, prospects have never been better for improvement.  In
addition to a tighter tie with Cepheids through the \sna\ comparison mentioned above, SBF
is ideally suited for calibration via the TRGB method. We document our initial SBF-TRGB
calibration in Appendix~\ref{sbf:trgb} of this work, which shows  excellent
agreement between  SBF and TRGB distances for two giant ellipticals in Virgo.  The
impending launch of the \textit{James Webb Space Telescope} (\jwst) will enable numerous
more TRGB distances to Virgo ellipticals in only modest amounts of observing time.

The systematic uncertainty in the TRGB absolute magnitude is  approaching 2\%
(see Appendix~\ref{sbf:trgb}).  If it can be reduced further through a combination of 
different approaches to the same 1.3\% level as for Cepheids, that 
would open the door to a 2\% constraint on the SBF zero point from a sample of 15
galaxies having both high-quality TRGB and SBF distances.  This estimate assumes 
realistic errors for both methods.  A single medium-sized \jwst\ proposal could achieve
this goal and give rise to a fully independent TRGB-SBF precision ladder, competitive with the
Cepheid-\sna\ ladder, for testing the significance of the discrepancy between local
and CMB-based measurements of \ho.  We look forward to ascending that ladder.

\vspace{-0.4cm}

\acknowledgments
We are grateful to Gabe Brammer for assistance with processing of the raw WFC3/IR data,
Brent Tully for guidance with the CF3 velocity data, and
Adam Riess, Dan Scolnic, and John Tonry for helpful conversations.
Support for this work was provided by NASA through grants HST-GO-14219, HST-GO-14654, and
HST GO-15265 from the Space Telescope Science Institute, which is operated by AURA, Inc.,
under NASA contract NAS\,5-26555. 
The MASSIVE Survey is supported in part by NSF grants AST-1815417 and AST-1817100.
C.-P.M. acknowledges support from the Heising-Simons Foundation and the Miller Institute for
Basic Research in Science.
This work was also supported by the International Gemini Observatory, a program of NSF’s
NOIRLab, which is managed by the Association of Universities for Research in Astronomy 
under a cooperative agreement with the National Science Foundation, on behalf of the Gemini
partnership of Argentina, Brazil, Canada, Chile, the Republic of Korea, and the United States
of America.
This paper is dedicated to the memory of Justin Ross Dougherty, a dear friend we lost while this was being written;
he so loved the concept of appendices. 

\facilities{HST(WFC3/IR, ACS), Pan-STARRS}


\clearpage

\appendix
\vspace{-0.4cm}
\section{SBF Zero Point from Tip of the Red Giant Branch\label{sbf:trgb}}

The SBF method works best with evolved stellar populations in morphologically regular
early-type galaxies. Cepheids do not exist in such galaxies, but the TRGB method is
well-suited for measuring their distances, and there have been initial efforts to calibrate
the TRGB with \gaia\ \citep{soltis2020}. This is an extremely promising route
for calibrating SBF via a direct tie to a geometrically calibrated distance indicator.
Here we take a first step along this route.

\subsection{Absolute Magnitude of the TRGB}

Before tying SBF to the TRGB method, we must enforce that the galaxies all use the same TRGB
absolute magnitude calibration.  Numerous TRGB calibrations have been published recently;
we list them for reference  in Table~\ref{trgbcals}.
Some of the calibrations are quoted in the \hst\ F814W bandpass, \Mtrgbacs,
while others are in the standard Cousins $I$~band, \MtrgbI. To convert
between the two, we use the following transformation from \citet{Riess2016} with the TRGB
color assumed by \citet{Freedman2019}:
 \begin{eqnarray}
   \Mtrgbacs \;&=&\; \MtrgbI \,+\, 0.02 \,-\, 0.018\,(V - I) \,; \\[3pt]
   \Mtrgbacs \;&\approx&\; \MtrgbI \,-\, 0.009 \,.
 \end{eqnarray}

\noindent
Thus, $\MtrgbI\approx\Mtrgbacs+0.01$.  For ease of comparison, the fourth column of
Table~\ref{trgbcals} shows the value of \MtrgbI, rounded to the nearest hundredth, for each
of the recent TRGB calibrations. Several of these take their zero point from the LMC, using the
DEB distance of \citet{Piet2019}, as we did in adjusting the Cepheid zero point for SBF. In
order to keep the two SBF calibrations independent, we prefer to use a TRGB calibration 
based on another geometric method, the maser distance to NGC\,4258.  For this purpose, we
take a simple average of the recent results of \citet{Reid2019} and \citet{Jang2020} using this
approach.  Since these studies use the same basic distance and photometric data, averaging
them does not reduce the error, and we adopt
 \begin{equation}\label{eq:trgb}
   \MtrgbI \;=\; -4.02 \pm 0.05\; \mathrm{mag}\,.
 \end{equation}
 This is very close to the value derived by \cite{JangLee2017} and the reassessment by
 \citet{capozzi2020} combining the distance from \citet{Reid2019} with the photometric measurements of \cite{JangLee2017}.

 Ultimately, the best calibration of \MtrgbI\ will come from \gaia, and the recent
 result of \citet{soltis2020} in Table~\ref{trgbcals}, using \gaia\ Early Data Release~3
 (EDR3) parallaxes, is   promising. Thus far, however, it is based on only the single
 large globular cluster  $\omega\,$Cen, likely the remnant nucleus of a stripped dE
 \citep[e.g.,][]{Maj2000,Hilk2000}. 
\citet{capozzi2020} use the same $\omega\,$Cen photometric data but adopted the kinematic distance from \citet{Baumgardt2019}
based on \gaia\ EDR2 data.
 \citet{cerny2020} use \gaia\ EDR2 proper motion
 data to assign membership for a sample of 46 globular clusters and determine their relative distances
 from the horizontal branch; the zero point is set from the DEB distance to $\omega\,$Cen.
  We look forward to analysis of the TRGB in an expanded sample of globular clusters using
 \gaia\ parallaxes. This will also enable a determination of the intrinsic scatter in the
 TRGB absolute magnitude, an error term that is generally ignored.

 \subsection{Tying the SBF Method to the TRGB}

\citet{Blak2009} tabulate high-quality \hst/ACS SBF distances for 134 early-type galaxies
mainly in the Virgo and Fornax clusters, and \citet{Jensen2015} present similar quality
WFC3/IR SBF measurements for 16 of these galaxies. These samples enabled the calibration of
the color dependence of the SBF magnitude in the giant ellipticals that we
measure at large distances.  We wish to tie these calibration samples to the TRGB method in
order to improve the SBF zero point calibration.

Unfortunately, there are few TRGB distances to giant ellipticals. Two exceptions are
M60/NGC\,4649 \citep{LeeJang2017} and M87/NGC\,4486  \citep{bird2010}. While not ideal,
the giant, dusty merger remnant NGC\,1316 (Arp~154, Fornax~A) is another galaxy that has a
high-quality \hst-based SBF distance and a published TRGB distance \citep{hatt2018}.
Table~\ref{tab:trgb-sbf} displays the SBF and TRGB distances for the three galaxies.  The
SBF distances in this table are the homogeneous set from \citet{Blak2009}, adjusted to the
revised LMC distance based on DEBs (see Sec.~\ref{sec:zpt}). The TRGB distances have been
adjusted for consistency with our adopted calibration in Eq.\,(\ref{eq:trgb}). For
NGC\,1316, we use the revised TRGB distance given by \citet{Freedman2019} before adjusting
to our adopted calibration.

We note that \citet{Freedman2019}, with the goal of increasing the number of
TRGB-\sna\ calibrators, also tabulate a ``TRGB distance,'' with a purported precision of
0.05~mag, for NGC\,1404, a giant elliptical just 50~kpc from the cD NGC\,1399 in the
core of the Fornax cluster. We have excellent SBF distances for these two galaxies.
However, the quoted TRGB distance is actually an average of the TRGB distances to the 
spiral galaxy NGC\,1365 and the merger remnant NGC\,1316.  NGC\,1365 lies 420~kpc in
projection from the Fornax core, has an angular size larger than any of the Fornax
ellipticals, and has been suggested to lie significantly in the foreground \citep{saha1997}.
NGC\,1316 is projected 1.25~Mpc from the core; if the separation  along the line
of sight is the same, the offset would be 0.14~mag. There is no evidence that these two
late-type galaxies at substantial separations from the  core  have distance
moduli within 0.05~mag of NGC\,1404. Thus, we cannot use it as a TRGB calibrator.

Table~\ref{tab:trgb-sbf} reports the weighted average offsets between the SBF and TRGB
distances when using just M60 and M87 in Virgo, and when using NGC\,1316 as well. These
averages differ by 0.07~mag.  While this is in statistical agreement, given
that we have so few calibrators, we want to ensure that all individual measurements are reliable.
NGC\,1316 is an irregular galaxy with prominent dust features, H$\alpha$
filaments, tidal loops, and shells indicative of recent merging \citep{schweizer1980}.
Based on the ages of the centrally concentrated population of bright, metal-rich star
clusters, the galaxy is believed to have undergone a major merger, with an associated
starburst, 2~to 3~Gyr ago \citep{goud2001,sesto2018}.
The IR SBF magnitude for NGC\,1316 is 0.25~mag brighter than expected based on the color
calibration defined by more evolved early-type galaxies.  This is consistent with predictions
from stellar population models if there is a 3~Gyr old population present
\citep{Jensen2015}.  In contrast, optical SBF distances are less affected by
younger populations because their effect is to move the galaxy both brighter
and bluer along the SBF-color relation, rather than simply scattering the SBF magnitude
brighter.  For this reason, we have more confidence in the optical SBF distance (shown in
Table~\ref{tab:trgb-sbf}) for this galaxy than the IR  one.

At present, the star formation in NGC\,1316 is mainly occurring at larger radius. Based on the mid-IR
emission, \citet{Temi2005} suggest that ``young, luminous stars are forming in the infalling
dusty gas, raising the dust temperature sufficiently to emit at 15~$\mu$m.''
Thus, unlike in spiral galaxies, where the star formation occurs in orderly fashion in the disk
while the halo hosts only ancient metal-poor stars, NGC\,1316 appears to have a
significant population of young stars at large radius.  Notably, the field used for
the TRGB measurement was about 10\arcmin\ from the galaxy's center, in an area of intense radio
emission \citep{ekers1983}.  We suspect that a dispersed 
population of asymptotic giant branch (AGB) stars associated with the prominent 3~Gyr, and
perhaps younger, component may have biased the TRGB measurement.  More dramatic biases have
occurred in TRGB estimates for more recent mergers at roughly the same distance
\citep{schweizer2008}. The young component at large radius in NGC\,1316 could account for
the observed difference in the TRGB and optical SBF distances for this galaxy.

In short, we lack confidence in NGC\,1316 as a reliable calibrator.  We therefore adopt the
mean offset of $-0.007\pm0.080$~mag between the TRGB and SBF distances, determined from the
two giant ellipticals in Virgo.  Including the 0.03~mag error in the tie between the optical
and IR SBF distances and the 0.05~mag error in the TRGB calibration, the final offset and
uncertainty in the SBF zero point from this TRGB comparison is  $-0.007\pm0.099$~mag, which
we round to $-0.01\pm0.10$~mag in Sec.\,\ref{sec:trgb} of the present work.

Ideally, we would make a direct tie between geometrically calibrated TRGB distances to giant
ellipticals and the IR SBF measurements to the same galaxies, avoiding the 0.03~mag error incurred
from the tie through the more extensive optical SBF data set.  However, at present, M60 is
the only IR SBF calibrator from \citet{Jensen2015} with a TRGB distance.  The result would
be similar, but with a larger error. Sec.~\ref{sec:future} presents a plan for improving
this situation.

\begin{deluxetable*}{lcCCCCCl}
\tablecaption{Ten Recent TRGB Absolute Calibrations\label{trgbcals}}
\tabletypesize{\small}
\tablehead{
 \colhead{} &
 \colhead{}      &
 \colhead{\phn$M_{\mathrm{Band}}^{\mathrm{TRGB}}$} &
 \colhead{\phn$M_I^{\mathrm{TRGB}}$} &
 \colhead{ $\sigma_{\textrm{tot}}$} &
 \colhead{ $\sigma_{\textrm{stat}}$} &
 \colhead{ $\sigma_{\textrm{sys}}$} &
 \colhead{ }\\[-6pt]
 \colhead{Reference \phantom{123}} &
 \colhead{Band} &
 \colhead{(mag)} &
 \colhead{(mag)} &
 \colhead{(mag)} &
 \colhead{(mag)} &
 \colhead{(mag)} &
 \colhead{Anchoring method \phantom{for zeropoint}}\\[-7pt]
 \colhead{(1)\phantom{1111}} &
 \colhead{(2)} &
 \colhead{(3)} &
 \colhead{(4)} &
 \colhead{(5)} &
 \colhead{(6)} &
 \colhead{(7)} &
 \colhead{(8)\phantom{111111111111}}
}
\startdata
\citet{Freedman2019}& F814W & -4.049 & -4.04 & 0.045 & 0.022 & 0.039 & DEB distance to LMC\tablenotemark{\small a} \\
\citet{Yuan2019}    & F814W & -3.970 & -3.96 & 0.046 & 0.038 & 0.026 & DEB distance to LMC\tablenotemark{\small b} \\ 
\citet{Freedman2020}& $I$   & -4.047 & -4.05 & 0.045 & 0.022 & 0.039 & DEB distance to LMC\tablenotemark{\small a} \\
\citet{soltis2020}  & $I$   & -3.961 & -3.96 & 0.040 & 0.011 & 0.038 & DEB distance to LMC\tablenotemark{\small c} \\
\citet{Reid2019}    & F814W & -4.012 & -4.00 & 0.044 & 0.030 & 0.032 & Maser distance to NGC\,4258  \\
\citet{Jang2020}    & F814W & -4.050 & -4.04 & 0.056 & 0.028 & 0.048 & Maser distance to NGC\,4258  \\
\citet{capozzi2020} & $I$   & -4.027 & -4.03 & 0.055 & 0.045 & 0.032 &  Maser distance to NGC\,4258  \\
\citet{capozzi2020} & $I$   & -3.960 & -3.96 & 0.067 & 0.064 & 0.021 & GAIA EDR2 kinematic $d$ to $\omega\,$Cen \\
\citet{soltis2020}  & $I$   & -3.970 & -3.97 & 0.062 & 0.041 & 0.047 & GAIA EDR3 parallax $d$ to $\omega\,$Cen \\
\citet{cerny2020}   & $I$   & -4.056 & -4.06 &  0.10\phn  & 0.022 & 0.101  & HB for 46 GCs + DEB in $\omega\,$Cen\tablenotemark{\small d} \\
%
%
\enddata
\vspace{-2pt}
\tablecomments{Columns list:  (1)~Calibration paper;
  (2)~reference band used in the study (Vega-based calibrations);
  (3)~derived TRGB absolute magnitude in reference band;
  (4)~absolute TRGB magnitude in standard Cousins $I$, assuming where needed $I =
  m_\mathrm{814W}+0.009$, and rounded to nearest hundredth;
  (5)~total error quoted from study, or quadrature sum of quoted random and
  systematic errors;
  (6)~quoted statistical error, or derived from information provided;
  (7)~quoted systematic error, or derived from information provided;
  (8)~distance method used for anchoring the zero point. 
 }
\vspace{-2pt}
\tablenotetext{\small a}{Extinction determined from observed TRGB color differences.}
\vspace{-2pt}
\tablenotetext{\small b}{Extinction  from \citet{haschke2011} OGLE reddening map.}
\vspace{-2pt}
\tablenotetext{\small c}{Extinction  from \citet{Skowron2020} OGLE reddening map.}
\vspace{-2pt}
\tablenotetext{\small d}{``HB'' refers to the horizontal branch, used by \citet{cerny2020} to shift
  the 46 globular clusters (GCs) into agreement before setting the distance zero point based on
  a DEB in $\omega\,$Cen \citep{Thompson2001}.}
\vspace{-0.75cm}
\end{deluxetable*}

\begin{deluxetable*}{lcccCCCl}
\tablewidth{0pt}   
\tablecaption{Homogenized SBF-TRGB Distance Comparisons\label{tab:trgb-sbf}}
\tabletypesize{\footnotesize}
\tablehead{\colhead{} &
\colhead{$(m{-}M)_{\mathrm{SBF}}$\tablenotemark{\footnotesize{a}}} &
\colhead{$\sigma_{\mathrm{SBF}}$} &
\colhead{$(m{-}M)_{\mathrm{TRGB}}$\tablenotemark{\footnotesize{b}}} &
\colhead{$\sigma_{\mathrm{TRGB}}$} &
\colhead{$\Delta(m{-}M)$\tablenotemark{\footnotesize{c}}} &
\colhead{$\sigma_\Delta$\tablenotemark{\footnotesize{c}}} &
\colhead{}\\[-8pt]
 \colhead{Galaxy} &
 \colhead{(mag)} &
 \colhead{(mag)} &
 \colhead{(mag)} &
 \colhead{(mag)} &
 \colhead{(mag)} &
 \colhead{(mag)} &
\colhead{Reference for TRGB\phantom{111111111}}
}\decimals
\startdata
NGC4486/M87   & 31.088 & 0.079 &  31.09 & 0.10 & -0.002 & 0.128 & \citet{bird2010}\\
NGC4649/M60   & 31.059 & 0.076 &  31.07 & 0.07 & -0.011 & 0.103 & \citet{LeeJang2017}\\
NGC1316       & 31.583 & 0.073 &  31.44 & 0.04 & +0.143 & 0.083 & \citet{hatt2018,Freedman2019}\\[2pt]
\hline  & \multicolumn{6}{c}{weighted average for Virgo galaxies:
   $\langle\Delta\mM\rangle \,=\; -0.007 \pm 0.080$} & \\[2pt]
\hline & \multicolumn{6}{c}{weighted average all three galaxies:
  $\langle\Delta\mM\rangle \;=\; +0.065 \pm 0.058$} & \\[2pt]
\hline
\enddata
\vspace{-1pt}
\tablenotetext{a}{SBF distance moduli from \citet{Blak2009}, reduced by 0.023~mag as described
  in Sec~\ref{sec:zpt}; $\sigma_{\mathrm{SBF}}$ is the statistical error as published.}
\vspace{-2pt}
 \tablenotetext{b}{TRGB distance moduli from references in last column, corrected by $-0.03$, $+0.02$, and $-0.02$~mag
    (M87, M60, and NGC\,1316, respectively) for consistency with our adopted zero point of
    $M_I^{\mathrm{TRGB}}=-4.02$ mag ($M_{814}^{\mathrm{TRGB}}=-4.03$ mag), which is an average of two recent
    TRGB calibrations based on the NGC\,4258 maser distance \citep{Reid2019,Jang2020}.
    The statistical errors $\sigma_{\mathrm{TRGB}}$ are as published; unlike the SBF errors,
    they include no allowance for intrinsic scatter in the absolute magnitude of the standardized candle.}
 \vspace{-2pt}
 \tablenotetext{c}{Difference in distance moduli: $(m{-}M)_{\mathrm{SBF}}-(m{-}M)_{\mathrm{TRGB}}$,
  and error in this difference {$\sigma_\Delta$}. }
\vspace{-0.5cm}
\end{deluxetable*}

\clearpage

\bibliographystyle{aasjournal}
\bibliography{main}
\end{document}